\documentclass{article}
\usepackage{amssymb,cite,multirow} 
\def\1ad{\mbox{\normalsize $^1$}}
\def\2ad{\mbox{\normalsize $^2$}}
\def\3ad{\mbox{\normalsize $^3$}}
\def\4ad{\mbox{\normalsize $^4$}}
\def\5ad{\mbox{\normalsize $^5$}}
\def\6ad{\mbox{\normalsize $^6$}}
\def\7ad{\mbox{\normalsize $^7$}}
\def\8ad{\mbox{\normalsize $^8$}}

\makeatletter
\@addtoreset{equation}{section}
\makeatother

\def\beq{\begin{equation}}                     %
\def\eeq{\end{equation}}                       %
\def\bea{\begin{eqnarray}}                     
\def\eea{\end{eqnarray}}                       
\def\nn{\nonumber} 
 
\def\cy {Calabi--Yau}

\def\clss {Clifford(6,6)} 
 
\def\0 {\nonumber} 
 
\def\bi{\bar \imath}
 
\def\bj{\bar \jmath}

\newcommand{\sla}{\slash\!\!\!}

\begin{document}

\setcounter{page}{0}
\begin{titlepage}
\titlepage
\vskip 5cm
\centerline{{ \bf \Large Type II Strings and}}
\vskip .5cm
\centerline{{ \bf \Large Generalized Calabi-Yau Manifolds}}
\vskip 1.5cm
\centerline{Mariana Gra{\~n}a$^{a,b}$, Ruben
Minasian$^a$, Michela Petrini$^{a,c}$
and Alessandro Tomasiello$^a$}
\begin{center}
\em $^a$Centre de Physique Th{\'e}orique, Ecole
Polytechnique
\\91128 Palaiseau Cedex, France\\
\vskip .4cm
$^b$Laboratoire de Physique Th{\'e}orique, Ecole Normale
Sup{\'e}rieure\\
24, Rue Lhomond 75231 Paris Cedex 05, France\\
\vskip .4cm
$^c$Laboratoire de Math{\'e}matiques et Physique Th{\'e}orique,
 Universit{\'e} Fran{\c c}ois Rabelais\\ 
Parc de Grandmont 37200, Tours, France

\vskip 2cm
\end{center}
\begin{abstract}

\vskip1cm
We show that  the supersymmetry transformations for type II string theories on
six-manifolds can be written as differential conditions on a pair of pure
spinors,  the exponentiated K{\"a}hler form $e^{iJ}$ and the holomorphic form
$\Omega$. The equations are explicitly symmetric under exchange of the
two pure spinors and a choice of even or odd-rank RR field. 
This is mirror symmetry for
manifolds with torsion. Moreover, RR fluxes  affect only one of the two
equations: $e^{iJ}$ is closed under the action of the twisted exterior
derivative in IIA theory, and similarly  $\Omega$ is closed in IIB. This means that
supersymmetric SU(3)-structure manifolds are always complex in IIB while they are twisted symplectic
in IIA. Modulo a
different action of the $B$--field, these are all generalized Calabi-Yau manifolds, as
defined by Hitchin. 

\end{abstract}
\vskip 0.5\baselineskip

\vfill
\begin{flushleft}
\end{flushleft}
\end{titlepage}

\section{Introduction}

Compactifications with fluxes have received much attention recently due 
to a number of interesting features. In many ways these can be seen as 
extensions of the more conventional compactifications on Ricci-flat 
manifolds. On the other hand, many aspects of the latter, most 
notably in the case of Calabi-Yau manifolds, still have to find their 
generalized counterparts.  
Mirror symmetry has been one of the most prominent and useful 
features of Calabi-Yau compactifications, and the question of its 
extension to compactifications with fluxes is both of conceptual and 
of practical interest. 

The issue of extending mirror symmetry to compactifications with fluxes 
has been studied recently in \cite{vafa,kstt,glmw,fmt,bb}. 
A first question is of course
within which class of manifolds this symmetry should be defined. A natural
proposal comes from the formalism of G--structures, recently used in many
contexts of compactifications with fluxes.
As shown in \cite{glmw,fmt}, 
mirror symmetry can be defined on manifolds of SU(3)
structure, thus generalizing the usual Calabi--Yau case. One of the points
which makes this symmetry non--trivial is that, as expected, 
geometry and NS flux mix in the transformation. On the contrary, RR 
fluxes are mapped 
by mirror symmetry into RR fluxes and  their transformation is
well-understood. However, for many reasons it would be better to have a 
formalism
that would incorporate geometrical data and fluxes in 
a natural way.
As a step forward in that direction, we propose to use  pure
  spinors of Clifford(6,6) as a formalism to describe SU(3)--structure compactifications. 
 
As far as we are concerned in
this introduction, \clss\ spinors are simply formal sums of
  forms, in analogy with  usual spinors, which are often realized as 
formal sums
of $(p,0)$ forms. A spinor is called 
 pure if it is annihilated by half of the gamma matrices. A pure spinor
 defines an SU(3,3) structure on the bundle $T + T^*$ on the manifold.  
If the spinor is also closed, the manifold is called by Hitchin \cite{hitchin}
a generalized \cy.

For a SU(3) structure on $T$, there are two pure spinors, $\varphi_1$ abd $\varphi_2$ which are 
orthogonal and of unit norm. An SU(3) structure is defined by a 
two--form $J$ and a three--form  
$\Omega$ obeying $J\wedge \Omega=0$ and $i\Omega\wedge\bar\Omega = (2J)^3/3!$. 
Then, the two pure spinors are $e^{i J}$ and $\Omega$.
We will show  that  
supersymmetry equations imply
differential equations for the  pure spinors, which are, schematically
\bea
\label{eq:pss}
e^{-f_1} d (e^{f_1}\varphi_1)&=&  H\bullet \varphi_1 \nn \\
e^{-f_2} d (e^{f_2} \varphi_2)&=&  H\bullet \varphi_2 + (F, \varphi) \ .
\eea
The operator $H\bullet$ is a certain action of the three--form $H$, involving
contractions and wedges but different from $H\wedge$.
So, both in IIA and IIB there is a ``preferred" pure spinor
(of the same parity as $F$ - namely $e^{iJ}$ in IIA and $\Omega$ in IIB) 
which does
not receive any back reaction from the RR fluxes, i.e. which is ``twisted'' closed. 
Then supersymmetry implies that 6-dimensional manifolds are all ``twisted'' generalized \cy\ \cite{hitchin}. The twisting
refers to the presence of the $H$ field. In the mathematical literature (and
in some physical applications \cite{rw})
this twisting is actually always appearing in the form $(d + H\wedge)$. It is
interesting to see that in general the inclusion of RR fluxes requires a
different form of twisting than the one usually assumed. Understanding the
origin of this twisting from first principles remains an important open problem.

\section{SU(3) structure and torsion versus fluxes}

We start by briefly introducing the  
notions of SU(3)-structure and intrinsic torsion with the help of which  
we will describe the non-Ricci-flat geometries under consideration. For a more
extensive review, see for example \cite{glmw}
and references therein.
A manifold with 
SU(3)-structure has all the group-theoretical features of a \cy, 
namely invariant two- and three forms, $J$ and $\Omega$ respectively. 
On a manifold of SU(3) holonomy, not only $J$ and $\Omega$ 
are well defined (nowhere vanishing, SU(3) invariant), but they are also 
closed: $dJ=0=d\Omega$. If they are not 
closed, 
$dJ$ and $d\Omega$ give a good measure of how far the manifold is from having 
SU(3) holonomy. Decomposing $dJ$ and $d \Omega$ in the different $SU(3)$ 
representations, we can write 
\begin{equation}
\label{eq:djdo}
\begin{array}{c}\vspace{.3cm} 
dJ = -\frac{3}{2}\, {\rm Im}(W_1 \bar{\Omega}) + W_4 \wedge J + W_3 \, ,\\
d\Omega = W_1 J^2 + W_2 \wedge J + \bar{W_5} \wedge \Omega\ \, .
\end{array}
\end{equation}
The $W$'s are the $(3 \oplus \bar{3} \oplus 1) \otimes (3 \oplus \bar{3})$
components of the intrinsic torsion: 
$W_1$ is a complex zero--form in $1 \oplus \bar 1$, $W_2$ is a complex primitive 
two--form, so it lies in $8 \oplus 8$, $W_3$ is a real primitive 
$(2,1) \oplus (1,2)$ form and it lies in 
 $6 \oplus \bar{6}$,  $W_4$ is a real one--form in $3 \oplus \bar{3}$, and 
finally $W_5$ is a complex $(1,0)$--form (notice that in (\ref{eq:djdo}) the  
$(0,1)$ part drops out), so its degrees of freedom are again  
$3 \oplus \bar{3}$.  
These $W_i$ allow to classify the differential type of any SU(3) structure. 

An SU(3) structure can 
be
defined also by a spinor $\eta$, which is nowhere vanishing, SU(3) invariant,
but not covariantly constant, unless the manifold has SU(3) holonomy. 
In terms of this, $J$ and $\Omega$ above are
defined as bilinears:
\bea \label{defJOmega}
\eta^\dagger \gamma_{mn}\gamma\,\eta &=& i J_{mn} \nn \\ 
-i \eta^\dagger \gamma_{mnp}(1+\gamma)\eta &=& \Omega_{mnp} \ .
\eea 
Torsion is induced by flux, so in any solution to the equations of motion
any nonvanishing torsion  has to be compensated by a nonvanishing flux
in the same representation. We can gain a lot of insight just by 
decomposing the fluxes into the different SU(3) representations, and
searching for any missing one. The following table shows the
number of times each representation appears for torsion, NS and RR fluxes:\\
\begin{center}
\begin{tabular}{|c|c|c|c|c|}  \hline 
 & $1\oplus \bar 1$ & $3\oplus \bar 3$ & $ \   \ \  \ \ 6\oplus \bar 6   \  \ \ \ \ $ & $8\oplus8 $ \\ \hline
Torsion & $1 \, (W_1)$ & $2 \, (W_4, W_5)$ & $1 \, (W_3)$ & $ 1 \, (W_2)$ \\ \hline
$H_3$ & $1$ & $1$ & $1$ & $0$ \\ \hline
IIA: $F_{2n}$ & $\ 2 \, (F_0,F_2,F_4)\  $ & $2 \, (F_2,F_4)$ & $0 $ & $\ 1 \, (F_2, F_4) \ $ \\ \hline
IIB: $F_{2n+1}$ & $1 \, (F_3) $ & $\ 3 \, (F_1,F_3,F_5)\ $ & $1 \, (F_3) $ & $0$ \\ \hline
\end{tabular}\\
Table 1: Decomposition of torsion and fluxes into SU(3) representations.
\end{center}

Just by looking at the table, we realize that in IIB there is no flux capable of
compensating the torsion class $W_2$. Thus, we can conclude that in any IIB solution,
$W_2$, which is an obstruction for getting complex geometry, has to vanish. In IIA 
there is no RR flux capable of compensating $W_3$ so, if this last torsion is not zero, it
must be compensated by NS flux. This means that in IIA there should be a relation $W_3 \sim H^{(6)}$ (the
$6$ denotes the representation). $W_3$ appears in the derivative of $J$, so it is an obstruction
to have symplectic geometry. Another torsion class, $W_1$, appears in both
$dJ$ and $d\Omega$, and represents an obstruction to have either complex or symplectic geometries.
If additionally $W_1=0$, which is true in any
IIA and IIB supersymmetric solution with SU(3) structure\footnote{This statement cannot be 
concluded just by looking at representations, since both in IIA and IIB there are enough scalars
in the flux to compensate $W_1$. It is derived by looking at all supersymmetry equations, as done
in \cite{paper}.}, then supersymmetric 6-manifolds with SU(3) structure are always complex in IIB        
while they are ``twisted symplectic'' in IIA (twisting refers to H-flux in the relation
$W_3 \sim dJ \sim  H^{6}$, we will expand on this later). 

Since IIA is related to symplectic geometries while IIB is associated to complex ones,
one immediately wonders if there is a mathematical construction that contains, or even more, extends,
both. That mathematical construction is generalized complex geometry. It has been
introduced by Hitchin \cite{hitchin} (see \cite{gualtieri} for details and further developments), and
recently used in string theory related context by \cite{Huybrechts,Kapustin,lmtz,KL}.   
It is clear that this formalism must be useful for mirror symmetry: although for the physical string
mirror symmetry is an exchange of \cy s, for the topological string it can be
formulated as sending symplectic manifolds into complex ones, and vice versa.

\section{Generalized complex geometry}

Usual complex geometry deals with the tangent bundle of a manifold $T$,
whose sections are vectors $X$, and separatedly, with the cotangent bundle
$T^*$, whose sections are 1-forms $\zeta$.  
In generalized complex geometry one deals with 
the direct sum of the tangent and cotangent bundle, $T \oplus T^*$
rather than the tangent (or cotangent) bundle itself, whose sections
are the sum of a vector field plus a one-form $X + \zeta$. 
The  standard machinery of complex geometry can be generalized to
this bundle.

To start with, let us consider the almost complex structure. 
If ordinary almost complex structures $J$ are bundle maps from $T$
to itself that square to $-\Bbb I_d$ ($d$ is the real dimension of the manifold), 
generalized almost complex structures
${\mathcal J}$ are maps of $T \oplus  T^*$ to itself that square to $-\Bbb I_{2d}$. 
As for an almost complex structure, ${\mathcal J}$ must also satisfy the 
hermiticity condition ${\mathcal J}^{t} {\mathcal I} {\mathcal J} = {\mathcal I}$, 
with the respect to the natural metric on $T \oplus  T^*$, 
${\mathcal I}={{0 \ \ 1} \choose {1 \ \ 0}}$. Such generalized almost complex structures
have the form
\beq \label{calJ}
{\mathcal J} = 
\left(
\begin{array}{cc}
J& P\\ L & K 
\end{array}
\right) \, ,
\eeq
where $J : T{M} \rightarrow T{M}$,  
$P : T^*{M} \rightarrow T{ M}$,  
$L : T{ M} \rightarrow T^*{ M}$ and  
$K : T^*{ M} \rightarrow T^*{ M}$.

The condition ${\mathcal J}^{t} {\mathcal I} {\mathcal J} = {\mathcal I}$ leads to  $K=-J^t$, $P=-P^t$ and
$L=-L^t$, so the matrix (\ref{calJ}) reads
\beq
 {\mathcal J} = 
\left(
\begin{array}{cc}
J& P\\ L & -J^t 
\end{array}
\right) \, ,
\eeq
with $P$ and $L$ antisymmetric matrices. The condition ${\mathcal J}^2=-\Bbb I_{2d}$
imposes further constrains for $J,P$ and $L$, in particular $J^2+PL=-\Bbb I_d$. 
From this, it is easy to see that usual complex structures are
naturally embedded into ${\mathcal J}$: they correspond to the choice 
\beq
{\mathcal J}_1 \equiv 
\left(
\begin{array}{cc}
J & 0 \\ 0 & -J^t 
\end{array}
\right) \, ,
\eeq
with $J_m\,^{n}$ an almost complex structure (i.e. $J^2=-\Bbb I_d$). 
Another example of generalized almost complex structure can be built
using a non degenerate 
two--form $\omega$, 
\beq
{\mathcal J}_2\equiv
\left(
\begin{array}{cc}
0 & -\omega^{-1} \\\omega & 0 
\end{array}
\right) \, .
\eeq

Given an almost complex structure one can build holomorphic and antiholomorphic 
projectors $\pi_{\pm}=\frac12 (1_{d}\pm iJ)$. Correspondingly, projectors can be  
build out of a generalized almost complex structure, $\Pi_{\pm}=\frac12 (1_{2d}\pm i {\mathcal J})$.
There is an
integrability condition for generalized almost complex
structures, analogous to the
integrability condition for usual almost complex structures.
For the usual complex structures integrability, namely the vanishing of the 
Nijenhuis tensor, 
can be written as the condition $\pi_{\mp} [\pi_{\pm} X,\pi_{\pm}Y]=0$, i.e. the
Lie bracket of two holomorphic vectors should again be holomorphic.  
For generalized almost complex structures, integrability condition reads
exactely the same, with $\pi$ and $X$ replaced by $\Pi$ and $X+\zeta$, and the Lie bracket
replaced by  
certain bracket on $T \oplus T^*$, called Courant bracket\footnote{The Courant
bracket is defined as follows: $[X+ \zeta , Y + \eta]_C=[X,Y] + {\mathcal L}_X \eta 
- {\mathcal L}_Y \zeta - \frac12 d(\iota_X \eta - \iota_Y \zeta)$.}. 
This bracket does not satisfy Jacobi identity in
general, but it does on the $i$--eigenspaces of ${\mathcal J}$.
In case
these new conditions are fulfilled, we can drop the ``almost'' and speak of
generalized complex structures. 

For the two examples of generalized almost complex structure
given above, ${\mathcal J}_1$ and ${\mathcal J}_2$, integrability condition
turns into a condition on the building blocks $J$ and $\omega$. For ${\mathcal J}_1$,
integrability of the 
generalized almost complex structure turns into 
the condition of $J$ being integrable as an almost complex structure
in $T$, thus making it a complex structure.
For ${\mathcal J}_2$, which was built from a two-form $\omega$, the condition 
becomes $d\omega=0$, thus making $\omega$ into a symplectic form.

These two examples are not exhaustive, and the most general
generalized complex structure interpolates between 
complex and symplectic manifolds. A generalized 
complex manifold is locally equivalent to the product
$\Bbb C^k \times (\Bbb R^{d-2k},\omega)$, 
where $\omega=dx^{2k+1} \wedge dx^{2k+2} + ... + dx^{d-1} \wedge dx^d$ is
the standard symplectic structure and $k\le d/2$ is called rank, and can be constant
or vary
over the manifold.

\subsection{Pure spinors in generalized complex geometry}

There is an algebraic correspondence between generalized almost complex
structures and  pure spinors of Clifford(6,6). In string theory,
the picture of generalized almost complex structures emerges naturally from
the worldsheet point of view \cite{lmtz}, while that of pure
spinors arises from the space-time side. Since it it this last
approach that we deal with, let us first review the formalism of
Clifford(6,6) spinors, and then show how to use pure
spinors in the context of generalized complex geometry. 

Spinors on $T$ transform under Clifford(6), whose algebra is $\{\gamma^m,\gamma^n\}=2 g^{mn}$.
There is a representation of this algebra in terms of forms. Using holomorphic
and anitholomorphic indices, we can take $\gamma^i=dz^i \wedge$, $\gamma^{\bi} g^{\bi j} \iota_j$.
\footnote{\label{iota} $\iota_{n}$: $\Lambda^p T^* \rightarrow
\Lambda^{p-1} T^*$, $\iota_{n} dx^{i_1} \wedge ... \wedge dx^{i_p} = p \delta^{[i_1}_n dx^{i_2} \wedge ... \wedge dx^{i_p]}$.} 
The (3,0)-form $\Omega$ can be used as a Clifford vacuum to construct a basis of spinors.
$\Omega$ is a pure spinor of Clifford(6), which means that it is anihilated by half
of the gamma matrices ($\gamma^i \Omega=0$). Acting with the rest of the
gamma matrices $\gamma^{\bi} $, $\gamma^{\bi \bj}$ and $\gamma^{\bi \bj \bar k}$, 
we can construct a basis of ``spinors'' made out
of (p,0)-forms. So Clifford(6) spinors are equivalent to (p,0)-forms. 

A similar story can be done with Clifford(6,6). To start with, there are twice
the number of generators as in Clifford(6), i.e. twelve. These are given by matrices
$\lambda^m, \rho_n$ obeying
$$
\{ \lambda^m, \lambda^n\} =0\ , \qquad 
\{ \lambda^m, \rho_n\} = \delta^m_n \ , \qquad
\{ \rho_m, \rho_n \}=0\ .
$$
We have chosen two different symbols, $\lambda$ and $\rho$, instead of the 
more commonly used $\gamma^m$ and $\gamma_m$,  to emphasize that these matrices
are independent, they cannot be obtained from 
each other by raising and lowering
indices with the metric.  
The representation of this algebra in terms of forms which is usually
taken, and to which we will stick, is $\lambda^m= dx^m
\wedge$, and $\rho_n = \iota_{n}$. $\Omega$ is still a good vacuum
of Clifford(6,6), as it is anihilated by $\lambda^i$ and $\rho_{\bi}$, which are
half of the gamma matrices, thus making it  
a pure spinor. Acting with the other half,  $\lambda^{\bi}$ and $\rho_{i}$we get forms of all possible degrees.
So Clifford(6.6) spinors are equivalent to (p,q)-forms. 

On a space with SU(3) structure on $T$, there two invariant forms, namely
$\Omega$ and $J$. $\Omega$ is a pure spinors, but $J$ is not. What is a pure
spinor instead is $e^{iJ} \equiv 1 + i J - \frac12 J\wedge J -\frac{i}{6} J\wedge J \wedge J$. It is
annihilated by $\rho_m + i J_{mn} \lambda^n$, as it is easy to check using
$J_m\,^n J_n\,^p = -\delta_m \,^p$. Thus on a space of SU(3) structure there are always
two pure spinors, $\Omega$ and $e^{iJ}$. It is shown in \cite{fmt} 
that the action of mirror symmetry  for manifolds with $SU(3)$ structure
that are $T^3$ fibrations over a 3-dimensional base is
\beq \label{mirror}
e^{iJ} \leftrightarrow \Omega \ .
\eeq
Furthermore, \cite{fmt} conjectured that this is the action
of mirror symmetry for any manifold with $SU(3)$ structure. By this proposal, mirror symmetry
is the exchange of two pure spinors.

There is a one to one correspondance between a generalized almost complex structure ${\mathcal J}$ and
a pure spinor $\varphi$. The six-dimensional space that anihilates the pure spinor 
should be equal to the $+i$ eigenspace of the generalized almost complex structure
that it is mapped to. Integrability condition for the generalized complex structure corresponds
on the pure spinor side to the condition \\
\begin{center}
${\mathcal J}$ is integrable $\Leftrightarrow $ $\exists $ vector $v$ and 1-form $\zeta$ such that $d\varphi = (v \llcorner + \zeta \wedge) \varphi$ 
\end{center} 
. \\
A generalized Calabi-Yau, as defined by Hitchin \cite{hitchin}, is a manifold on which a closed pure spinor exists.

There is also a possibility of adding a three--form $H$ to the story. 
Using a three--form, the Courant bracket can be modified\footnote{$[X+\zeta, Y+\eta]_H=[X+\zeta,Y+\eta]_C + \iota_X \iota_Y H$.}, and with it 
the integrability
condition. Not surprisingly, this corresponds also to a modification of the 
condition on the pure spinor, which now becomes 
\begin{equation}
  \label{eq:Hw}
(d + H\wedge) \varphi=(v\llcorner + \xi ) \varphi\   
\end{equation}
If we decompose $\varphi$ in forms, $\sum \varphi_{(k)}$, the
condition means that $d \varphi_{(k)} + H\wedge \varphi_{(k-2)}= v\llcorner \varphi_{(k+2)} + \zeta \wedge \varphi_{(k)}$
for every $k$. 

\subsection{Supersymmetry equations for pure spinors}

In this section we will use the supersymmetry
equations in type IIA and type IIB supergravity to derive equations on the two pure spinors. 
The equations we derive do not encode all the
information coming from the supersymmetry conditions. They are rather the
counterpart of the internal gravitino, in that they encode derivatives of
$J$ and $\Omega$, which come from covariant deriavtives of the spinor
in the original internal gravitino equation. They capture the information
about the intrinsic torsion of the manifold; but in general from supersymmetry 
there are more conditions arising, equaling components of fluxes (and derivatives of
the dilaton and warping) among each other. These conditions are explicitly given in \cite{paper}. 
 
To get equations for the pure spinors one starts with the internal gravitino equation which,
in the democratic formulation of \cite{demo} reads
\beq 
\label{eq:susyg} 
\delta \psi_m = D_m \epsilon + \frac{1}{4} H_m {\mathcal P} \epsilon + \frac{1}{16} e^{\phi}  
\sum_n  \sla \! {F_{2n}} \, \Gamma_{m} {\mathcal P}_n \, \epsilon  \, ,
\eeq 
where $F_{2n}=dC_{2n-1} - H \wedge C_{2n-3}$ are the modified RR field strengths with
non standard Bianchi identities, that we will
call from now on simply RR field strengths; $n=0, \ldots,5$ for IIA and  
$n=1/2, \ldots,9/2$ for IIB and $H_M \equiv \frac{1}{2}H_{MNP} \Gamma^{NP}$
and ${\mathcal P} = \Gamma_{11}$,  
${\mathcal P}_n = \Gamma_{11}^n \sigma^1$ for IIA,
while ${\mathcal P} = -\sigma^3$, ${\mathcal P}_n = \sigma^1$ for $n+1/2$ even and 
${\mathcal P}_n = i \sigma^2$ for $n+1/2$ odd for IIB.
The two Majorana-Weyl supersymmetry parameters of type II supergravity are
arranged in the doublet  $\epsilon= (\epsilon_1,\epsilon_2)$.

The ``total" RR field 
involves both the  field strengths and their duals, and a self-duality 
relation is still to be imposed
\bea
F_{2n}  
= (- 1)^{Int[n]} \star_{10} F_{10-2n} \, .
\label{sd10}
\eea
In order to preserve 
4d Poincare invariance, RR fluxes should be of the form 
\beq 
\label{eq:rrfs} 
F_{2n} = {\hat F}_{2n} + Vol_4 \wedge {\tilde F}_{2n-4} \, .
\eeq 
Here ${\hat F}_{2n}$ stands  for purely internal fluxes. 
The self-duality of $F_{2n}$, Eq.(\ref{sd10}) becomes ${\tilde F}_{2n-4} = 
(-1)^{Int[n]} \star_6  {\hat F}_{10-2n}$, and allows to write the  
RR part of (\ref{eq:susyg})  in terms of the internal fluxes only.
From now on we will work only with internal fluxes, and drop the hats in $F$.  

The ten dimensional Majorana-Weyl spinors $\epsilon_1,\epsilon_2$, which have opposite chirality
in IIA and the same chirality in IIB, can be decomposed
\begin{eqnarray} 
\epsilon_1 & = &  \zeta_{+} \otimes \eta^1_{+} + \zeta_{-} \otimes 
\eta^1_{-} \, ,
\nonumber\\ 
\epsilon_2 & = & \zeta_{+} \otimes \eta^2_{-}  
+ \zeta_{-} \otimes \eta^2_{+} \, , 
\label{IIAans} 
\end{eqnarray} 
in IIA, where $\zeta$ and $\eta^i$ are chiral spinors in 4 and 6 dimensions,
respectively. The Majorana condition implies also
$(\zeta_{+})^* =  \zeta_{-}$, $(\eta^i_{+})^{*} = \eta^i_{-}$. 
For IIB, the two spinors can be decomposed
\beq 
\epsilon_i = \zeta_{+} \otimes \eta^i_{+} + \zeta_{-} \otimes \eta^i_{-} \, . 
\eeq 
On a manifold of SU(3) structure there is only one nowhere vanishing invariant spinor, $\eta$.
So $\eta_1$ and $\eta_2$ should be related to $\eta$, which also means that $\epsilon_1$
and $\epsilon_2$ are related, as should be the case for ${\mathcal N}=1$ supersymmetry.
We write the relation as
\beq \label{defab}
\eta_+^1=a \eta_+ \ , \ \ \  \eta_+^2=b \eta_+
\eeq
In supersymmetry equations, we will use the combinations
\beq \label{defalphabeta}
\alpha \equiv a+ ib \ , \ \ \  \beta \equiv a-ib \ .
\eeq
Coming back to the pure spinors, the strategy to get equations for them is to use the fact that
 we can map a form
(or a formal sum of them) to an element of the usual Clifford algebra, 
Clifford(6):
\begin{equation}
  \label{eq:clmap}
C\equiv\sum_k \frac{1}{k!}C^{(k)}_{i_1\ldots i_k} dx^{i_i}\wedge\ldots\wedge dx^{i_k}\qquad
\longleftrightarrow\qquad
\sla C \equiv
\sum_k \frac{1}{k!}C^{(k)}_{i_1\ldots i_k} \gamma^{i_i\ldots i_k}_{\alpha\beta} \ .
\end{equation}
An object in Clifford(6) can also be seen as a bispinor, since it has two free
spinor indices. So we have realized \clss\ spinors as bispinors, which are
more useful in string theory.  Another useful technical fact is that one can realize
$\lambda$ and $\rho$ also as combinations of the more familiar $\gamma$'s 
acting on the left and on the right of a bispinor. For example, 
$\lambda^m C^{(k)} \longleftrightarrow$ 
$\frac{1}{2}(\gamma^m {\sla C^{(k)}} \pm {\sla C^{(k)}} \gamma^m)$ 
when the plus (minus) sign corresponds to  $k$ even (odd).

A crucial fact is that  $e^{i J}\!\!\!\!\! 
\begin{picture}(10,10)
\put(0,0){\line(1,1){10}}
\end{picture}
$
and $\sla \Omega$ can be reexpressed in terms of tensor products of $\eta$. Using 
Fierz rearrrangement, one can show 
\beq 
\eta_{\pm} \otimes \eta^{\dagger}_+ = 
\frac{1}{4} \sum_{k=0}^6 \frac{1}{k!}  
\eta^{\dagger}_+ \gamma_{i_1 ... i_k} \eta_{\pm} 
\gamma^{i_k ... i_1} 
\label{eq:fz}
\eeq  
Using the expression for $J$ and $\Omega$ in terms of $\eta$, Eq.(\ref{defJOmega}),
it is possible
to express the pure spinors as tensor products of the standard spinor 
defining the SU(3) structure
\bea \label{Fierz} 
\eta_{\pm} \otimes \eta^{\dagger}_{\pm} &=& 
\frac{1}{8} 
e^{\mp i J}\!\!\!\!\! 
\begin{picture}(10,10)
\put(0,0){\line(1,2){5}}
\end{picture} \, ,\nn \\ 
\eta_{+} \otimes \eta^{\dagger}_{-} &=& 
-\frac{i}{8} \sla {\Omega} \, ,\nn \\ 
\eta_{-} \otimes \eta^{\dagger}_{+} &=& 
-\frac{i}{8} \sla \overline{\Omega} \, .
\eea 
where the extra factor of $1/2$ with respect to (\ref{Fierz})  comes from the normalization chosen for the spinors,
$\eta^{\dagger}_{\pm} \eta_{\pm} = \frac12$. 
Then, the exterior derivative  $d(e^{-iJ})$ can be re-expressed in
the bispinor picture as the anticommutator 
$$
\{ \gamma^m, D_m(\eta_+\otimes \eta_+^\dagger) \} \  .
$$
The covariant derivative here is meant to be a bispinor covariant derivative, 
which corresponds to the ordinary covariant derivative of forms under the 
Clifford map, and which anyway reduces to exterior derivative when we fully 
antisymmetrize, as usual. To compute this object, one can use Leibniz rule 
for the covariant derivative of the bispinor, reducing it to 
$\{ \gamma^m, D_m(\eta_+)\otimes \eta_+\}$ plus its complex conjugate.
Using the internal gravitino equation (\ref{eq:susyg}) for the covariant derivative 
of the spinor, gives
\begin{eqnarray*}
\mathrm{IIA:}&& - [\sla \partial (2A -\phi +\log \alpha)
  +\frac\beta{4\alpha}\sla H]
\eta_+\otimes\eta_+^\dagger - (\partial_m \alpha +\frac\beta{4\alpha}H_m)
\eta_+\otimes\eta_+^\dagger \gamma^m \, ,\\
\mathrm{IIB:}&& - [\sla \partial (2A -\phi +\log \alpha)
  -\frac\beta{4\alpha}\sla H ]
\eta_+\otimes\eta_+^\dagger 
- (\partial_m \alpha -\frac\beta{4\alpha}H_m + \\
&& \hspace{7cm} -\frac i{4\alpha}e^{\phi} \sla F_{B}\gamma_m)
\eta_+\otimes\eta_+^\dagger \gamma^m \, ,
\end{eqnarray*}   
where $\alpha$ and $\beta$ are defined in (\ref{defalphabeta}),
$A$ is the warp factor, i.e. the metric has the form \\
$$
ds^2= e^{2A} (\eta_{\mu \nu} dx^{\mu} dx^{\nu}) + ds_6^2
$$
and $F_B = \alpha F_1 - \beta F_3 + \alpha F_5$ is a sum of IIB RR fluxes.  
Notice that in IIA $F$ has disappeared. This is because it would have
multiplied $\gamma_m \eta_-\otimes \eta_+^\dagger \gamma^m$. This
expression is zero because $\eta_-\otimes \eta_+^\dagger=-\frac
i8\sla\bar\Omega$,  
and $\gamma_m \gamma^{npq} \gamma^m=0$ in
six dimensions. This technical circumstance is what allows us to
make $F$ disappear in one of the pure spinor equations for both IIA
and IIB. It is now only required to go from the bispinor
picture  back to the form picture, inverting the Clifford 
map (\ref{eq:clmap}). The equations we obtain are the following.
For type IIA we have
\bea
e^{-f} d \Big(e^f e^{i J} \Big) &=&  -\frac{1}{2}
\frac{Re(\alpha\bar\beta)}{|\alpha|^2 + |\beta|^2} H\bullet e^{iJ} \, ,
\label{p1}\\
e^{-g} d \Big(e^g \Omega\Big) &=& 
-\frac14\frac{\beta^2+\alpha^2}{2\alpha{}\beta{}} H\bullet \Omega + \label{p2} \\
&-& \frac{e^\phi{}}{16}\frac1{2\alpha\beta} \Big( F \cdot
(-\frac14 e^{-i J} +1 + i\mathrm{vol}) - (-\frac14 e^{i J} +1 -i \mathrm{vol}) \cdot F^* \Big) \,
, \nonumber
\eea  
and in type IIB 
\bea
e^{-f} d \Big(e^f e^{iJ}\Big) &=& 
\frac12 \frac{Re(\alpha\bar\beta)}{|\alpha|^2 + |\beta|^2} H\bullet
e^{iJ} +
\label{p3} \\
&-& i\frac{ e^\phi{}}{16} \frac1{|\alpha{}|^2 + |\beta{}|^2}\Big( F\cdot  
(-\frac14e^{-i J} + 1 +i \mathrm {vol}) - (-2e^{-i J} + 1 +i \mathrm{vol}) \cdot F 
\Big) \ , \nonumber \\
e^{-g} d \Big(e^g \Omega \Big) &=&  \frac14\frac{\beta^2+\alpha^2}{2\alpha\beta{}} 
H\bullet \Omega{} \label{p4}\, .
\eea 
In both cases $f=2A-\phi+\log(|\alpha|^2 + |\beta|^2)$ and $g=2A-\phi +
\log(\alpha\beta)$, and $F \equiv (|\alpha|^2 - |\beta|^2)F_+ +
(\alpha\bar\beta-\bar\alpha \beta)F_-$, where $F_+$ is the hermitian
piece of the RR total form ($F_+=F_0+F_4$ in IIA, $F_+=F_1 +F_5$ in IIB) and
$F_-$ is the antihermitian piece ($F_-=F_2 +F_6$ in IIA and $F_-=F_5$ in
IIB). A dot $\cdot$ indicates the Clifford product between forms\footnote{$F \cdot G$ is obtained by first building the bispinor 
$\not F \not G $ and then using the map (\ref{eq:clmap}) to get back
the corresponding form.} and vol is the volume form.
 The operator $H\bullet$ is the same for all equations
and is defined by
\begin{equation}
  \label{eq:Hop}
H\bullet \equiv  H_{mnp} \left(dx^m dx^n \iota^p -\frac13
  \iota^m \iota^n \iota^p \right)\ .
\end{equation}

Although the RR piece is not very nice, it has a similar form in both
cases too. Most importantly, the action of the NS sector is always the
same. 

Given the mathematical
discussion, it is natural to wonder if the operator $H\bullet$ we found has a
realization in terms of a twisting of the Courant bracket. This remains as an open problem.
Note however that the combination $d + H\bullet$ does not square to zero, unlike
$d + H \wedge$. 

With this caveat (or technical clarification) in mind, we will 
call any action of $H$-flux a twisting. The main outcome of the equations (\ref{p1}--\ref{p2})  for IIA and 
(\ref{p3}--\ref{p4})  for IIB is that in each case there is
one pure spinor equation that contains an exterior derivative and
$H$-twist. Thus, having a twisted closed pure spinor, or in other words
twisted generalized Calabi-Yau, is a necessary condition for having an
${\mathcal N} = 1$ vacuum. All the backgrounds with SU(3) structure
constructed so far satisfy this
condition.

The pure spinor that is twisted close in each case has the same parity as
the RR-flux: even for IIA ($e^{iJ}$) and odd for IIB ($\Omega$).
This respects the mirror symmetry exchange (\ref{mirror}).

The condition $e^{iJ}$ being twisted closed in IIA means that in IIA
supersymmetric manifolds are twisted symplectic. In IIB, on the contrary,
$\Omega$ is twisted closed. Decomposing 
(\ref{}) order by order, one gets $H \llcorner \Omega=0$ 
(so $H$ does not contribute to $W_1$), and $d\Omega$ is (3,1) ($H$ does not
-and cannot- contribuite to $W_2$). So supersymmetric manifolds with SU(3) structure
in IIB are always complex.

\section{Discussion}

To summarize, we obtained that supersymmetry implies that 
the 6-dimensional compactification manifolds of type II 
with $SU(3)$ structure are always twisted generalized \cy s.
This means that they have one twisted closed pure spinor,
$e^{iJ}$ for IIA and $\Omega$ for IIB,
which has the same parity as the RR-flux. 
Twisting refers to the action of the 3-form $H$, $d_H = d + H \bullet$ 
(see (\ref{eq:Hop}), which works
differently than the way considered by \cite{hitchin}, $d_H= d + H \wedge$.  
Understanding the supergravity twisting from first principles 
remains an open problem.  

There are quite a few other open problems related to generalized \cy ``compactifications''.
One is the issue about global tadpoles: what kind of compact manifolds are suitable, i.e. evade no-go theorems? 
In the case type IIB on warped-\cy s, which are a particular case
of generalized \cy, O3 planes give the appropiate negative tension and RR-charge source to cancel
tadpoles. For other kind of generalized \cy s, which are supersymmetric given a set of fluxes
the orientifold planes needed to cancel global tadpoles break supersymmetry (for example, 
in IIB solutions corresponding to bound states of D3 and D5-branes, there is no known
combination of O3 and O5-planes that preserves supersymmetry).

Another key open question is the deformation problem for twisted
operators (while the discussion of $H\wedge$ twisting started as early as
in \cite{rw} and is still far from being complete, as mentioned
the very origin of $H\bullet$ is yet to be understood). It seems very
likely that the generalized complex geometry provides the right framework
for these problems, and  the understanding of the moduli spaces of
the generalized \cy s, and consequently the string spectra in
flux compactifications will hopefully be achieved soon.

\section*{Acknowledgements} We would like to thank Andrew Frey, Jan Louis, Simon Salamon,
Maxim Zabzine and Alberto Zaffaroni for useful discussions. M.G. would like to thank the organizers of
Strings 2004 for the invitation to present this work.

This work is supported in part by EU contract
HPRN-CT-2000-00122 and by INTAS contracts 55-1-590 and 00-0334.
MG was partially supported by European Commission Marie Curie Postdoctoral
Fellowship under contract number MEIF-CT-2003-501485.


\begin{thebibliography}{00}




\bibitem{paper}
M.~Gra\~na, R.~Minasian, M.~Petrini and A.~Tomasiello,
``Supersymmetric backgrounds from generalized Calabi-Yau manifolds,''
arXiv:hep-th/0406137.

\bibitem{vafa}
C.~Vafa,
``Superstrings and topological strings at large N,''
J.\ Math.\ Phys.\  {\bf 42} (2001) 2798
[arXiv:hep-th/0008142].

\bibitem{kstt}
S.~Kachru, M.~B.~Schulz, P.~K.~Tripathy and S.~P.~Trivedi,
``New supersymmetric string compactifications,''
JHEP {\bf 0303}, 061 (2003)
[arXiv:hep-th/0211182].

\bibitem{glmw}S.~Gurrieri, J.~Louis, A.~Micu and D.~Waldram,
``Mirror symmetry in generalized Calabi-Yau compactifications,'' Nucl.\
Phys.\ B {\bf 654}, 61 (2003) [arXiv:hep-th/0211102].

  
\bibitem{fmt}S.~Fidanza, R.~Minasian and A.~Tomasiello,
``Mirror symmetric SU(3)-structure manifolds with NS fluxes,''
arXiv:hep-th/0311122.

\bibitem{bb}
O. ~Ben-Bassat, 
``Mirror symmetry and generalized complex manifolds,''
arXiv:math.aG/0405303.


\bibitem{hitchin}
N.~Hitchin,
``Generalized Calabi-Yau manifolds,''
arXiv:math.dg/0209099.


\bibitem{rw}
R.~Rohm and E.~Witten,
``The Antisymmetric Tensor Field In Superstring Theory,''
Annals Phys.\  {\bf 170} (1986) 454.




\bibitem{gualtieri}
M.~Gualtieri,
``Generalized complex geometry,''
Oxford University DPhil thesis, arXiv:math.DG/0401221.

\bibitem{Huybrechts}
D.~Huybrechts,
``Generalized Calabi-Yau structures, K3 surfaces, and B-fields,''
arXiv:math.ag/0306162.

\bibitem{Kapustin}
A.~Kapustin,
``Topological strings on noncommutative manifolds,''
arXiv:hep-th/0310057.



\bibitem{lmtz}
U.~Lindstrom, R.~Minasian, A.~Tomasiello and M.~Zabzine,
``Generalized complex manifolds and supersymmetry,''
arXiv:hep-th/0405085.

\bibitem{KL}
A.~Kapustin and Y.~Li,
``Topological sigma-models with H-flux and twisted generalized complex
manifolds,''
arXiv:hep-th/0407249.

\bibitem{demo}
E.~Bergshoeff, R.~Kallosh, T.~Ortin, D.~Roest and A.~Van Proeyen,
``New formulations of D = 10 supersymmetry and D8 - O8 domain walls,''
Class.\ Quant.\ Grav.\  {\bf 18} (2001) 3359
[arXiv:hep-th/0103233].



\end{thebibliography}
\end{document}